\documentclass{nature}
\usepackage{amsmath,amssymb}
\usepackage{lineno}
\usepackage{graphics,graphicx}
\usepackage{adjustbox}
\usepackage{color}
\usepackage{longtable}
\newcommand{\mnras}{{\it Mon. Not. Roy. Astron. Soc.\, }}
\newcommand{\aap}{{\it Astron. Astrophys.\, }}
\newcommand{\apjl}{{\it Astrophys. J. Lett.\, }}
\newcommand{\apj}{{\it Astrophys. J.\, }}
\newcommand{\nat}{{\it Nature\, }}
\newcommand{\pasj}{{\it Publ. Astron. Soc. Japan\, }}
\newcommand{\gcn}{{\it GCN Circ.\, }}

\title{Simultaneous Radio and Optical Polarimetry of GRB 191221B Afterglow}

\author{Yuji Urata$^{1,2}$, Kenji Toma$^{3,4}$, Stefano Covino$^5$, Klaas Wiersema$^{6,7}$, Kuiyun Huang$^8$, Jiro Shimoda$^9$, Asuka Kuwata$^4$, Sota Nagao$^4$, Keiichi Asada$^{10}$, Hiroshi Nagai$^{11,12}$, Satoko Takahashi$^{13,14,12}$, Chao-En Chung$^1$, Glen Petitpas$^{15}$, Kazutaka Yamaoka$^{16, 17}$, Luca Izzo$^{18}$,
 Johan Fynbo$^{19}$,  Antonio de Ugarte Postigo$^{20}$,  Maryam Arabsalmani$^{21,22}$,  \& Makoto Tashiro$^{23,24}$}

\begin{document}
\maketitle

\begin{affiliations} 
\item Institute of Astronomy, National Central University, Chung-Li 32054, Taiwan
\item MITOS Science CO., LTD., New Taipei 235, Taiwan
\item Frontier Research Institute for Interdisciplinary Sciences, Tohoku University, Sendai 980-8578, Japan
\item Astronomical Institute, Graduate School of Science, Tohoku University, Sendai, 980- 8578, Japan 
\item Istituto Nazionale di Astrofisica / Brera Astronomical Observatory, via Bianchi 46, 23807 Merate (LC), Italy
\item Department of Physics, University of Warwick, Coventry CV4 7AL, United Kingdom
\item School of Physics and Astronomy, University of Leicester, University Road, Leicester LE1 7RH, United Kingdom
\item Center for General Education, Chung Yuan Christian University, Taoyuan 32023 Taiwan
\item Department of Physics, Graduate School of Science, Nagoya University, Furo-cho, Chikusa-ku, Nagoya 464-8602, Japan
\item Academia Sinica Institute of Astronomy and Astrophysics, Taipei 106, Taiwan
\item National Astronomical Observatory of Japan, 2-21-1 Osawa, Mitaka Tokyo 181-8588, Japan
\item Department of Astronomical Science, School of Physical Sciences, The Graduate University for Advanced Studies, SOKENDAI, 2-21-1 Osawa, Mitaka, Tokyo, 181-8588, Japan
\item Joint ALMA Observatory, Alonso de Cordova 3108, Vitacura, Santigao, Chile
\item NAOJ Chile Observatory, Alonso de Cordova 3788, Oficina 61B, Vitacura, Santiago, Chile
\item Harvard-Smithsonian Center for Astrophysics, 60 Garden Street, Cambridge, Massachusetts 02138, USA
\item Institute for Space-Earth Environmental Research (ISEE), Nagoya University, Furo-cho, Chikusa-ku, Nagoya, Aichi 464- 8601, Japan  
\item Division of Particle and Astrophysical Science, Graduate School of Science, Nagoya University, Furo-cho, Chikusa-ku, Nagoya, Aichi 464-8602, Japan
\item DARK, Niels Bohr Institute, University of Copenhagen, Jagtvej 128, 2200 Copenhagen, Denmark
\item The Cosmic Dawn Center Niels Bohr Institute Copenhagen University Jagtvej 128 DK-2200 Copenhagen N
\item Artemis, Observatorie de la C\^{o}te $D^,$Azur, Universite\'C\^{o}te $D^,$Azur, CNRS, Nice, 06300, France
\item Excellence Cluster ORIGINS, Boltzmannstra{\ss}e 2, 85748 Garching, Germany
\item Ludwig-Maximilians-Universit\"at, Schellingstra{\ss}e 4, 80799 M\"unchen, Germany
\item Department of Physics, Saitama University, Shimo-Okubo, Saitama, 338-8570, Japan
\item Institute of Space and Astronautical Science, JAXA, Sengen, Tsukuba, Ibaraki, Japan
\end{affiliations}

\begin{abstract}
Gamma-ray bursts (GRBs) are the most luminous transients in the universe and are utilized as probes of early stars, gravitational wave counterparts, and collisionless shock physics. In spite of studies on polarimetry of GRBs in individual wavelengths that characterized intriguing properties of prompt emission and afterglow, no coordinated multi-wavelength measurements have yet been performed. Here, we report the first coordinated simultaneous polarimetry in the optical and radio bands for the afterglow associated with the typical long GRB 191221B. Our observations successfully caught the radio emission, which is not affected by synchrotron self-absorption, and show that the emission is depolarized in the radio band compared to the optical one. 
Our simultaneous polarization angle measurement and temporal polarization monitoring indicate the existence of cool electrons that increase the estimate of jet kinetic energy by a factor of $>$ 4 for this GRB afterglow. Further coordinated multi-wavelength polarimetric campaigns would improve our understanding of the total jet energies and magnetic field configurations in the emission regions of various types of GRBs, which are required to comprehend the mass scales of their progenitor systems and the physics of collisionless shocks.
\end{abstract}

 GRB 191221B was detected on 21 December 2019, 20:39:13 UT, and its X-ray afterglow was rapidly identified by the Neil Gehrels Swift Observatory\cite{gcn26534}. 
  The optical afterglow was discovered by the MASTER auto-detection system\cite{gcn26537}. 
Optical polarization with a possible time evolution in the early afterglow phase was also detected by the Southern African Large Telescope (SALT) and Very Large Telescope (VLT)(Extended Data Table \ref{optpol})\cite{buckley21}.
 The redshift was measured as $z =1.148$ based on metal absorption lines in the optical afterglow observed by VLT/X-shooter\cite{gcn26553}. The isotropic equivalent energy of $E_{\gamma, {\rm iso}}$ = $(3.6 \pm 0.4)\times10^{53}$ erg and the rest-frame peak energy of the time-integrated spectrum $E^{\rm src}_{\rm peak}= 810 \pm 65$ keV were derived by the Konus-Wind observation (with the standard cosmological parameters $H_{0} = 67.3$ km/s/Mpc, $\Omega_{M}$ = 0.315, and $\Omega_{\Lambda}$ = 0.685)\cite{gcn26576}.
 The duration of the prompt emission in the 15-350 keV band is $48.0 \pm 16.0$ sec\cite{gcn26562}. These prompt emission properties obey the empirical {\bf $E^{\rm src}_{\rm peak}-E_{\gamma,{\rm iso}}$} correlation (Extended Data Figure \ref{amati}) and indicate that GRB 191221B is one of the typical long GRBs.

\begin{figure}
\centering
\includegraphics[width=18cm]{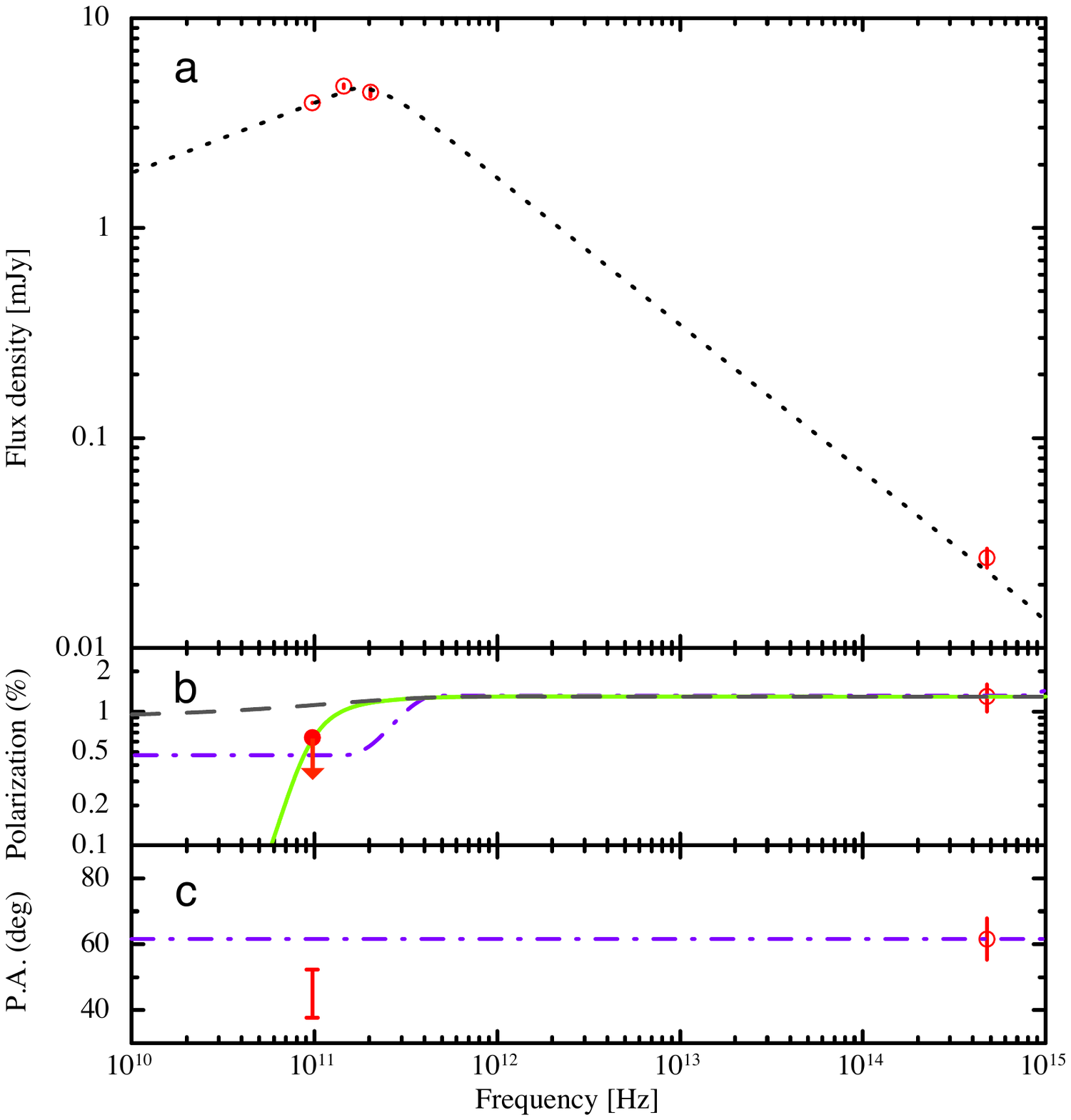}
\caption{{\bf Spectral flux distribution and polarization spectrum of GRB 191221B afterglow at 2.5 days.}  a, Spectral flux distribution (red points). The black dotted line is the forward shock model fit to the observed data. b, PDs at 97.5 GHz (3-$\sigma$ upper limit) and optical $R$ band (red points), and polarization spectrum of the simple one-zone model (grey dashed line), the plasma-scale magnetic field model (purple dashed-dot line) and the cool electron model (green solid line). c, PAs at 97.5 GHz (1-$\sigma$ range) and optical $R$ band (red points). The observed difference of PAs with $\sim$90\% confidence level (i.e., 16.6 $\pm$ 9.6 deg.) supports the cool electron model. 
The plasma-scale magnetic field model predicts a constant PA over the frequencies (e.g., purple dashed-dot line). All error bars represent 1-$\sigma$ uncertainties. 
}
\label{polspec}
\end{figure}

The first semi-simultaneous polarimetry for the afterglow between millimeter and optical bands was conducted at 2.5 days after the burst by using Atacama Large Millimetre/submillimetre Array (ALMA) and VLT (Figure \ref{polspec}). The VLT observation measured a linear polarization degree (PD) of 1.3 $\pm$ 0.2 \% (here, we employed the systematic errors of 0.1\% reported by \cite{vltpolsys} and the range with $3\sigma$ confidence level is 0.9$-$1.8\%) with a polarization angle (PA) of 61.6 $\pm$ 6.3 deg at the $R$ band. Hereafter, we noted 1-$\sigma$ errors for our measurements without special notification.
The low dust extinction and Serkwski law \cite{serkowski73} show an intrinsic origin of the polarization (see Methods, Extended Data Figures \ref{xshooter},\ref{galactic},\ref{xspec} and Extended Data Table \ref{xray}).
The PD is consistent with other optical afterglows (average of 1.6\% among 84 polarimetric measurements)\cite{covino16}.
The ALMA observation put the upper limit on PD of 0.6\% with $3\sigma$ confidence level at 97.5 GHz. The detection in Stokes-$U$ maps and non-detection in Stokes-$Q$ maps (Extended Data Figure \ref{stokesmap}) constrained the range of PA as 37.7$-$52.3 deg with $1\sigma$ confidence level. 
Therefore, this simultaneous polarimetry between optical and radio bands indicates depolarization in the radio band.
The significantly low upper limit is also consistent with the first detection of linear polarization in the GRB radio afterglow (i.e., 0.2\% for the low-luminosity GRB 171205A)\cite{urata19}.

\begin{figure}
\centering
\includegraphics[width=15cm]{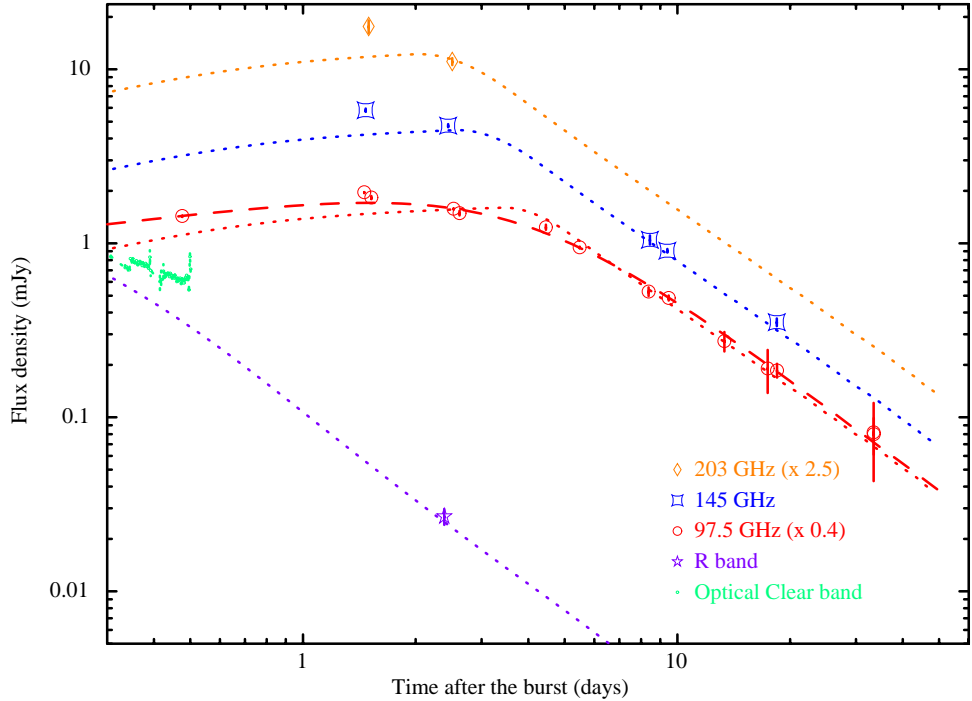}
\caption{{\bf Radio afterglow light curve of GRB 191221BA with the simultaneous optical ($R$ band) polarimetric observation.} The red dashed line indicates the best fitted smoothly connected broken power-law functions of the 97.5 GHz light curve. The radio light curves and the optical $R$ band photometric measurement are described by the standard forward shock synchrotron radiation model. Differences in the early optical afterglow (green small circles) and its wiggles may be caused by the magnitude-to-flux conversion of optical observations made by the very broad-band clear filter. The forward shock model describes the passing of synchrotron spectral peak over the ALMA observing band around 4 days, which is consistent with observed spectrum change between 2.5 and 9.5 days (Figure \ref{sed}). All error bars represent 1-$\sigma$ uncertainties. }
\label{lc-zoom}
\end{figure}

\begin{figure}
\includegraphics[width=15cm]{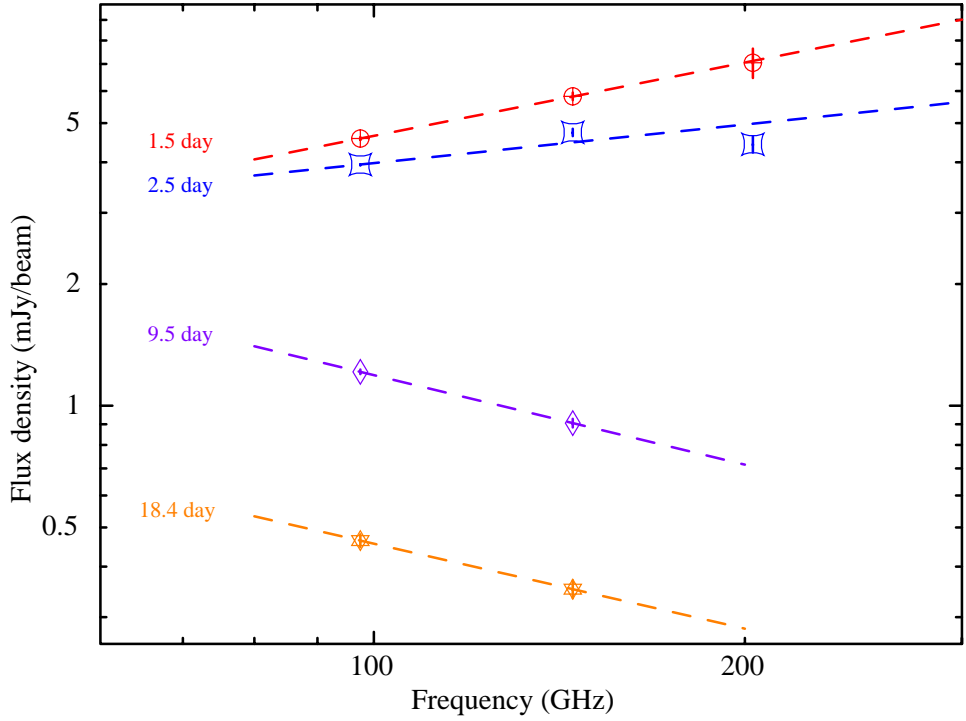}
\caption{
{\bf  Spectral flux distributions of the GRB 191221B afterglow at 1.5, 2.5, 9.5 and 18.4 days after GRB.} The photometry with high signal-to-noise characterized the spectral slope $\beta$ as $0.602\pm0.007$ at 1.5 days, $0.3\pm0.15$ at 2.5 days, and $-0.7$ at 9.5 and 18.4 days, respectively. The changing of spectral indices from positive to negative indicated the passing of the spectral peak frequency through the radio band. All error bars represent 1-$\sigma$ uncertainties.}
\label{sed}
\end{figure}

The synchrotron self-absorption (SSA) effect, which suppresses polarization below the SSA frequency ($\nu_{a}$), is not a reliable explanation for the observed depolarization. ALMA observations at 97.5 GHz, 145 GHz and 203 GHz (Figure \ref{lc-zoom}, Table \ref{almapol} and Extended Data Table \ref{tablephot}) show that the light curve at the 97.5 GHz band exhibited a broken power-law evolution with power-law indices of 
$\alpha = 0.26\pm 0.02$ (before the break) and 
$\alpha = -1.62\pm 0.13$ (after the break), and a break-time at 
$3.77\pm 0.35$ days, where and hereafter we describe the temporal and spectral properties of the flux density as $F_{\nu} \propto t^{\alpha}\nu^{\beta}$. The multi-frequency measurements (Figure \ref{sed}) showed that the spectral slope changed from positive power-law index of $\beta \sim 0.602\pm0.007$ at 1.5 day and $\beta \sim 0.32\pm0.15$ at 2.5 day to negative one ($\beta \sim -0.7$) at 9.5 and 18.4 day. These spectral slopes are in disagreement with the SSA effect which leads to $\beta=2$ \cite{sari98,zhang04}.

These afterglow properties are instead
reproduced by the standard model\cite{sari98,zhang04} of optically-thin synchrotron emission from an expanding shock in a uniform density medium with an isotropic energy that increases with time by long activity of the central engine $E_{\rm iso} = 9.4\times10^{52} (t/1\;{\rm day})^{0.25}$ erg, an ambient medium density $n = 5.9$ cm$^{-3}$, a fraction of shocked energy transferred to non-thermal electrons $\epsilon_{e} = 6.5 \times 10^{-2}$, that to magnetic field $\epsilon_{B} = 1.2\times10^{-2}$, the energy spectral slope of non-thermal electrons $p = 2.4$, the jet opening half-angle $\theta_{j} = 2.6$ deg, and the viewing angle $\theta_{v} =1.9$ deg (see Methods). The model lines in the top panel of Figures 1 and 2 are the results of our numerical calculations of synchrotron flux by taking account of the equal observed times of photons\cite{granot99,shimoda21}. The model also explains the X-ray and optical afterglows (Extended Data Figure \ref{lcall}). The peak frequency at $\sim$ 200 GHz in the top panel of Figure \ref{polspec} is the synchrotron frequency of minimum-energy electrons $\nu_{m}$, and the temporal and spectral changes around $t \sim 4$ days are found to be consistent with the crossing of $\nu_m$ at the observed frequencies.
The energy scale is comparable to the observed $\gamma$-ray energy $E_{\gamma, {\rm iso}}$, and the micro-physical parameters are also consistent with other cases\cite{panaitescu02}.
The deviation of the observed radio flux from the model light curve at 1.5 day (see Figure \ref{lc-zoom}) and its slightly hard spectrum (see Figure \ref{sed}) would imply additional emission component, but it is negligible at $t \geq 2.5\;$day (see Methods for more discussion).

If the magnetic field is ordered in the emitting shocked region, the PD of synchrotron emission is $\simeq 70\%$ at $\nu > \nu_m$ while $50\%$ at $\nu < \nu_m$, and the polarization direction is perpendicular to both the magnetic field direction and the line of sight in the comoving frame, where the electron momentum distribution is assumed to be isotropic\cite{rybicki79}. Usually, however, the magnetic field in the shocked region is tangled through its amplification process from the field of the ambient medium by some type of instability\cite{medvedev99,sironi07}, and therefore, the net polarization of observed synchrotron emission is reduced, depending on the magnetic field configuration in the visible region. One may consider a simple one-zone model in which the PDs at various frequencies, i.e., $\nu > \nu_m$ and $\nu < \nu_m$, are reduced by the same factor\cite{gruzinov99,sagiv04,toma08} (the grey dashed line in Figure \ref{polspec} b), but this model is ruled out by the observed PD data in the radio and optical bands.

One of the most actively discussed magnetic field amplification processes is Weibel instability, which occurs at relativistic collisionless shocks and generates strong magnetic fields with random directions on plasma skin depth scale\cite{medvedev99,gruzinov99,spitkovsky08,keshet09}. In this case, the field component parallel to the shock plane may be dominant. This anisotropy results in a sizable PD at each position of the shock, although the field is tangled on the tiny scale\cite{sari99,ghisellini99}. We numerically calculated the net linear polarization in various frequencies based on the synchrotron emission model explained above (see Methods for more details). As shown in the middle panel of Figure \ref{polspec}, the PD at $\nu \lesssim \nu_{m}$ is much lower than that at $\nu > \nu_{m}$ since the surface brightness distribution is significantly non-uniform at the frequencies $\nu > \nu_{m}$\cite{shimoda21,granot99}. 
This property can be consistent with the data. However, this model has a clear prediction that the PA at $\nu > \nu_{m}$ and $\nu \lesssim \nu_{m}$ are the same or 90 deg different\cite{shimoda21}.  
The difference in the observed PA at the radio and optical bands (the bottom panel of Figure \ref{polspec}) does not support this model. The temporal evolution of PD is also incompatible with this model (see the Extended Data Figure \ref{lcall}).

Another possible process of magnetic field amplification is magnetohydrodynamic instabilities at the shock. These include Richtmyer-Meshkov instability, which occurs in the case of ambient medium with inhomogeneous density\cite{sironi07,inoue13}. In this case, the magnetic field directions in the shocked region can be random mainly on hydrodynamic scales comparable to the typical width of the bright region in the shock downstream, and the internal Faraday depolarization can be significant in the radio band if the number of non-thermal electrons is a fraction $f (<1)$ of the total shocked electrons\cite{toma08}. The fraction $1-f$ of the total shocked electrons remain so cool that cause the Faraday depolarization on the emission from the non-thermal electrons. The true isotropic energy is $E_{\rm iso}/f$ and the true fraction of non-thermal electron energy is $\epsilon_e f$ in this case\cite{eichler05}. We calculate the PD in the one-zone model similar to Sokoloff et al. (1998)\cite{sokoloff98}, and plot the model in the middle panel of Figure \ref{polspec} (see more details in Methods). 
To explain the observed PDs, the Faraday rotation in the shocked region should be significant at $\nu < 100$ GHz. This leads to an upper limit $f \lesssim 0.3$. 
The difference of the surface brightness distribution in optical and radio bands or the contribution from the ordered magnetic field\cite{sokoloff98}  explain the observed difference in PAs.

 Our observations performed the first simultaneous polarimetry between the millimeter and optical bands for the typical long GRB 191221B.
 The multi-frequency observations of the afterglow were also described by the standard model.
 The measured radio PD was significantly lower than the optical one. 
 Two plausible models that provide new insight into GRB sciences were considered for the origin.  The measured PAs and the PD temporal evolution indicated the Faraday depolarization caused by cool electrons in the hydrodynamic-scale magnetic field. 
 Our observation consolidates a new methodology for revealing the 
 total jet energies and collisionless shock physics of various types of GRBs.
 If $f$ is very small for low-luminosity GRBs and/or short GRBs, their true total jet energies are much larger than the current estimates, which may increase their neutrino production rates\cite{murase06,kimura17} and the lower bound of total explosion energies or mass scales of progenitor systems.

\begin{methods}

\noindent {\bf Atacama Large Millimetre/submillimetre Array and Atacama Compact Array Observations}\\
A total of 11 epochs of radio observations were conducted using the Atacama Large Millimetre/submillimetre Array (ALMA) and Atacama Compact Array (Table \ref{almapol} and Extended Data Table \ref{tablephot}). Four epochs (0.5, 1.5, 2.5, and 9.5 days) of observations were performed with the polarization mode at 97.5 GHz (i.e., Band 3). Multi-frequency observations were managed with the photometry mode at 145 GHz among 5 of the 11 epochs. At 1.5 and 2.5 days, additional two photometry at 203 GHz were also conducted. Semi-simultaneous optical polarimetry was also performed 2.5 days after GRBs using the Very Large Telescope (VLT).

 Regarding the ALMA calibrations, the bandpass and flux were calibrated using observations of J1037-2934, and J1036-3744 was used for the phase calibration. The polarization calibration was performed by observing of J1058+0133. The raw data were reduced at the East Asian ALMA Regional Center (EA-ARC) using CASA (version 5.6.1)\cite{casa}. We further performed interactive CLEAN deconvolution imaging with self-calibration. The Stokes $I$, $Q$, and $U$ maps were CLEANed with an appropriate number of CLEAN iterations after the final round of self-calibration. The off-source RMS levels in $I, Q,$ and $U$ are consistent with the expectations for thermal noise alone. The quantities that can be derived from the polarization maps are the polarized intensity ($\sqrt{Q^2+U^2}$), polarization degree ($100\sqrt{Q^2+U^2}/I$ \%, PD), and polarization position angle ($1/2\arctan(U/Q)$, PA). By applying the polarization calibration to the phase calibrator J1036-3744 and creating Stokes maps for 6, 9, and 18 epochs during the 3-hr observing period, we confirm that the stability of linear polarization degree is $<$0.02 \%, which is consistent with the systematic linear polarization calibration uncertainty of 0.033\% for compact sources. 
The ACA data were flagged, calibrated, and imaged with standard procedures with CASA (version 5.6.1). The bandpass and flux were calibrated using observations of J1058+0133 and J1107-4449. The observations of J1018-3123 were used for the phase calibration.

\noindent{\bf Very Large Telescope Spectroscopic Observations}

The VLT also obtained X-shooter spectra for the afterglow of GRB\,191221B at $\sim 10$ and $\sim 34$\,hr after the GRB onset.
X-shooter spectra cover a very large wavelength range, from the UV atmospheric cutoff to more than 2\,$\mu$m. This range is covered by the three arms of the instrument: the UVB, VIS, and NIR arms.
Observations consisted of two sets of 600s exposures in the three arms using the AB nod-on-slit offset mode. Data in the UVB and VIS arms have been reduced using the stare mode standard data reduction, namely by extracting the science spectra as if they were obtained without any offset. The NIR arm was extracted using the standard X-shooter NOD mode pipeline. Each single spectrum has been corrected for slit-losses, due to the finite aperture of the X-shooter slit, and subtracted by residual sky emission. The final stacked spectrum has been finally corrected for telluric features, whose response correction for the VIS and NIR arms has been estimated from the spectrum of the standard telluric star \cite{goldoni2006,modigliani2010,selsing2019}.

A full study of these spectra is well beyond our interest, and in this work, our goal is just to model the afterglow-only spectrum (the one obtained at $\sim 10$\,hr after the burst) for deriving an estimate of the optical extinction along the line of sight. This would allow us to compute a plausible maximum level of host-galaxy dust-induced (i.e., non-intrinsic to the GRB afterglow) polarisation. Properly connecting the three X-shooter arms require a careful cross-calibration again beyond our interests, therefore we limited our analysis to the UVB arm covering the rest-frame wavelength range from approximately 1650 to 2550\AA\  (from 3450 to 5500\AA\  in the observer frame). The resolution of the X-shooter spectra is 0.2\AA /bin. We first rebinned the spectra to 20\AA /bin by the algorithm described in Carnall (2017)\cite{carnal2017} and then manually removed all the main emission or absorption lines. The resulting spectrum shows small scale variations, likely an artifact of the reduction process related to the Echelle spectrograph different orders. This does not affect our fits, although we had to add, in quadrature, a systematic error of $7.5 \times 10^{-18}$ erg s$^{-1}$ cm$^{-2}$ $\AA^{-1}$ to the uncertainties computed by the reduction pipeline.
We fit the afterglow spectrum in this wavelength range by a simple power-law 
 affected by rest-frame extinction following the Small or Large Magellanic Cloud (SMC and LMC) or the Milky Way (MW) extinction curves \cite{gordon2003,gordon2009}. The rest-frame extinction turns out to be very low, and therefore the three extinction recipes yield essentially the same results (see also \cite{covino2013}): 
$\beta = -0.97^{+0.14}_{-0.07}$,  E$_{\rm B-V} < 0.038$ (95\% upper limit). The best-fit for the SMC recipe is shown in the Extended Data Figure \ref{xshooter}.

The VLT also obtained spectro-polarimetric observations with the FORS2 instrument at about 10\,hr after the GRB onset. These data were already reported in Buckley et al. (2021)\cite{buckley21}. Data show (their Fig. 4) a fairly constant polarization level and position angle. In Buckley et al. (2021)\cite{buckley21} spectro-polarimetry obtained with the SALT/RSS telescope at $\sim$ 3\,hr after the burst is also reported. The more modest S/N prevents us further analyses on these data regarding the possible evidence for a Serkowski law behavior. We have downloaded the VLT spectrum and carried out a fit with a Serkowski law\cite{serkowski73} and the predictions for afterglow polarisation in the optical band (i.e., constant polarisation). As expected, both scenarios can provide an acceptable fit to the data, although for the Serkowski law the wavelength corresponding to the polarisation maximum is pushed in the far ultraviolet ($\sim 200$\,nm) in order to have a roughly constant polarisation in the wavelength range covered by the FORS2 spectro-polarimetry. This is a rather unusual result but not totally unprecedented \cite{patat2015}. However, the Serkowski law fit is not statistically favored compared to the afterglow only since it requires a larger number of free parameters. Therefore, also considering the low extinction along the line of sight derived by the analysis of the X-shooter spectra, an intrinsic origin (i.e., due to the afterglow) of the observed polarisation compared to the dust-induced hypothesis appears to be more in agreement with the data.

\noindent {\bf Very Large Telescope Polarimetric Observations}\\
Polarimetric observations were acquired using the Focal Reducer and low dispersion Spectrograph (FORS2) mounted on the VLT.  A Wollaston prism was inserted in the light path to split the image of each object in the field into two orthogonal polarization components. A mask was used to avoid overlap of the two  images; we used the FORS2 $R$ band filter. For each position angle $\phi$/2 of the half-wave plate rotator, we obtained two simultaneous images of cross-polarization at angles $\phi$ and $\phi +90^{\circ}$.
We obtained observations at position angles 0, 22.5, 45 and $67.5^\circ$ of the half-wave plate. This technique allowed us to remove any differences between the two optical paths (ordinary and extraordinary ray), including the effects of seeing and airmass changes. With the same setup we also observed polarized and unpolarized standard stars to convert position angles from the telescope to the celestial reference frame, and to correct for the small instrumental polarization introduced by the telescope.
Reduction, i.e., bias, flat-field correction, bad pixel masking, etc., were carried out following standard recipes. Aperture photometry was obtained for the target and several nearby sources in the field. 
We also confirmed that the GRB polarization measurement is unaffected by Galactic dust induced polarization (Extended Data Figure \ref{galactic}.)
We used custom software tools based on the python astropy library (http://www.astropython.org). More details on polarimetric data analysis are reported in Covino et al. (2003)\cite{covino03} and Wiersema et al. (2012)\cite{wiersema12}.

Extended Data Figure \ref{lcall} shows the temporal evolution of polarization combined with the optical polarimetric results reported by the Buckley et al. (2021)\cite{buckley21}. 
There are three epochs of optical observation including our measurements. 
Buckley et al. (2021) made two epochs of polarimetry during the optical light curve showed the wiggles and reported the marginal decreasing of polarization degree (by $\sim$0.3 \%) over a timescale of $\sim$7 h.Re
We derived the PDs and PAs in the wavelength range of $R$ band based on data reported by Buckley et al. (2021) (Extended Table \ref{optpol}).
Based on the two-sample t-test, 
the PA between radio and optical is different at $\sim$90\% confidence level.
The temporal evolution of PDs is also inconsistent with the plasma-scale magnetic field model. These properties do not support the plasma-scale magnetic field model.

\noindent{\bf X-ray spectrum of GRB afterglows with optical polarimetry}\\
We checked the hydrogen column density of the line of sight ($N_{H}$), which is one of the indicators of dust extinction of afterglows at their host galaxies.
There are 8 known-$z$ GRB afterglows, including GRB 191221B, available with optical polarimetry and {\it Swift} X-ray observations. 
For 6 events (GRB 080928, GRB 091018, GRB 091208B, GRB 110205A, GRB 121024A, and GRB 131030A), the optical observations reported the detection of intrinsic polarization \cite{covino16,wiersema12,uehara12,gorosabel11,wiersema14,king14}. 
For GRB 190114C, Jordana-Mitjans et al. (2020)\cite{jordana-mitjans20} reported the detection of polarization induced by the dust in the host galaxy. 
The X-ray data obtained by the {\it Swfit}/XRT were collected from the UK Swift Science Data Centre \cite{evans07,evans09}. We rebinned the spectra so that each spectral bin contains more than five counts. Using the software XSPEC 12, we perform spectral fitting with a single power law modified with intrinsic and Galactic absorptions, the latter of which are fixed at values calculated from Willingale et al. (2013)\cite{willingale13}. 
TBabs and zTBabs that incorporate three absorption elements (i.e., gas, molecules, and grains)\cite{wilms00} were used to describe the spectral absorptions.
The derived best-fitting values are summarized in Extended Data Table \ref{xray}.
Using Schady et al. (2010)\cite{schady10}, the measured $N_{H}$ are converted to $A_{V}$.
Five events, including GRB 191221B, exhibited the intrinsic absorption column density of 10 to the 21st power.
The intrinsic absorption column density of GRB 191221B is the smallest one ($N_{H}=1.6^{+0.9}_{-0.8}\times10^{21}$ ${\rm cm}^{-2}$). 
This result is consistent with the low dust extinction derived by the analysis of the VLT/X-shooter spectra.
In contrast, the GRB 190114C X-ray spectrum is highly obscured by the intrinsic absorption column density of $N_{H}=8.5\times10^{22}$ ${\rm cm}^{-2}$ (Extended Data Figure \ref{xspec}). These results naturally explain the dust-induced optical polarization of GRB 190114C and support the intrinsic polarization observed in other events.
Hence, these results also indicate the intrinsic origin of the optical polarization measured on the GRB 191221B optical afterglow.

\noindent {\bf Afterglow modeling}\\
The observed radio spectra and the light curves in the radio, optical and X-ray bands are explained by the standard forward shock model\cite{sari98,zhang04}. The temporal change of the spectral slope in the 97.5-203 GHz (Figure \ref{sed}) and the breaks of the 97.5 and 145 GHz light curves (Figure \ref{lc-zoom})at $t \sim 4$ days indicate the crossing of the synchrotron frequency of minimum-energy electrons $\nu_{m}$ at the observed frequencies. From the observed spectral slope $\beta \sim -0.7$ at $\nu >\nu_{m}$, the electron energy spectral index is estimated as $p =-2\beta+1 \sim 2.4$. Then the theoretical temporal decay indices at $\nu < \nu_{m}$ and at $\nu > \nu_{m}$ are 1/2 and $3(1-p)/4 \sim -1.1$, respectively, in the case in which the collimated forward shock expands in uniform density medium and its edge is not visible due to the strong relativistic beaming. These are not consistent with the observed indicies $\sim 0.26 ~(t \lesssim 4\;{\rm days})$ and $\sim -1.6 ~(t \gtrsim 4\;{\rm days})$. After the edge becomes visible (without sideways expansion of the shock), the geometrical flux reduction $\theta_{j}^2/(1/\Gamma)^2 \propto t^{-3/4}$ results in the decay indices $-1/4$ (at $\nu < \nu_{m}$) and $-1.8$ (at $\nu > \nu_{m}$), where $\theta_{j}$ and $\Gamma$ are the opening half angle and Lorentz factor of the shock.
The wind type ambient medium gives decay indices $-3/4$ (at $\nu < \nu_{m}$) and $-2.3$ (at $\nu > \nu_{m}$).
We find that the observed indices can be fit by the model in which the ambient density is uniform and the energy continues to be injected to the shock by long activity of the central engine\cite{rees98,zhang01}.
We note that our assumption of no sideways expansion of shock is based on the results of high-resolution hydrodynamic simulations, which show the collimated shock after the time of $1/\Gamma \sim \theta_{j}$ expands sideways logarithmically, not exponentially\cite{zhang09,vaneerten12}. 

We performed numerical calculations of flux from the shock with a fixed $\theta_j$ which evolves as the Blandford-McKee (BM) self-similar solution, by taking account of the equal arrival time surface of photons\cite{granot99,shimoda21}.
Then we tried to fit the model flux to the observed data by adjusting the model parameters, namely the isotropic energy $E_{\rm iso}$, the ambient medium density $n$, the fraction of shock energy carried by the electrons $\epsilon_{e}$, that carried by amplified magnetic field $\epsilon_{B}$, and the viewing angle $\theta_v$, as well as $\theta_j$, and $p$. 
This model can fit the data in the radio, optical and X-ray bands (Figures \ref{polspec}, \ref{lc-zoom}, and Extended Data Figure \ref{lcall}). The model parameters are constrained to be $E_{\rm iso} = 9.4\times10^{52} (t/1\;{\rm day})^{0.25}$ erg, $n =5.9 \;{\rm cm}^{-3}$, $\epsilon_{e} = 6.5\times10^{-2}$, $\epsilon_{B} = 1.2\times10^{-2}$, $\theta_v = 1.9\;$ deg, $\theta_j = 2.6\;$deg, and $p = 2.4$. These values of $n$, $\epsilon_{e}$, and $\epsilon_{B}$ are typical of GRB afterglows\cite{panaitescu02}.  

It is well known that X-ray flares with fast variability sometimes dominate the forward shock emission. If the flares have broad emission spectra, they might also affect the radio light curves. The slight deviations of radio data from the forward shock model light curves at $t=0.5$ and 1.5 days might be related with the X-ray flares observed at the similar times. Since they have fast variability, they may not contribute to the spectrum at $t=2.5$ days. Figure \ref{polspec} simply indicates that the standard forward shock synchrotron spectrum with the radio data at $t=2.5$ days (i.e. the peak flux $\sim 5$ mJy, $\nu_m \sim 200$ GHz, and the spectral index (at $\nu>\nu_m$) $\beta \sim -0.7$) can explain the optical data, and does not require any additional emission component.

The possibility that the short-lived reverse shock explains the radio emission at $t \gtrsim 1.5$ day is excluded since the minimum synchrotron frequency of the reverse shock\cite{gao13} $\nu_m^r \sim 200$ GHz with the peak flux $\sim 5$ mJy requires $\epsilon_e \sim 1$.
We also examined possible long-lasting reverse shock emission in the long-active central engine model like our model shown above. Suppose the reverse shock emission is dominant in the radio band while the forward shock is in the optical band, the difference in polarization in the two bands could be caused by possible difference in magnetic field structures in the two shocked regions. However, this scenario is also disfavored due to a high value of $\epsilon_e$. According to Sari \& M\'{e}s\'{a}ros (2000)\cite{sari00}, the minimum synchrotron frequencies of forward and reverse shocks for our model parameters at $t = 1.5$ days are $\nu_m \sim 9.8 \times 10^{11}$ Hz and $\nu_m^r \sim 2.9 \times 10^{10}$ Hz, respectively, where the equal arrival time surface of photons is not taken into account. To increase $\nu_m^r$ to a frequency at $\nu_m$, without changing the forward shock X-ray flux $\propto \epsilon_e^{p-1} E_{\rm iso}^{(p+2)/4}$, requires $\epsilon_e \sim 0.53$. 
This value is unusually high, compared to $\epsilon_e \lesssim 0.3$ estimated by systematic studies using well-sampled multi-frequency observations\cite{panaitescu02,yost03,cenko10}

\noindent {\bf Polarization in the plasma-scale magnetic field model}\\
The synchrotron polarization depends on the magnetic field configuration at each position in the shocked fluid. Here we focus on the turbulent magnetic field with coherence length on plasma skin depth scale, which is many orders of magnitude smaller than the shock width. Such magnetic field is created by Weibel instability at relativistic collisionless shocks\cite{medvedev99,gruzinov99,spitkovsky08,keshet09}, 
and in this case the field may be anisotropic, i.e., $\xi^{2} \equiv 2\langle B_{\parallel}^2 \rangle/\langle B_{\perp}^2 \rangle \neq 1$, where $\langle B_{\parallel}^2 \rangle$ and $\langle B_{\perp}^2 \rangle$ are the averages of the powers of magnetic field components parallel and perpendicular to the shock normal, respectively. Based on this model, we can calculate the local Stokes $Q$ and $U$ parameters corresponding to the surface brightness of the shock by averaging the emissivity with respect to the field directions at each position, and find that the synchrotron emission at each position is polarized due to the anisotropy of the turbulent magnetic field\cite{sari99,ghisellini99,rossi04}.
The polarization directions are symmetric around the line of sight, so that the net polarization degree is non-zero only when the visible region of angular size $\sim 1/\Gamma$ includes the jet edge and becomes asymmetric\cite{sari99,ghisellini99,rossi04}.

We numerically calculated the linear PDs in various frequencies based on the light curve model explained above. The parameter value $\xi^{2} = 0.56$ leads to the optical PD $\simeq 1.3\%$ at $t = 2.5\;$days. In the optical band, the surface brightness has a peak at $\theta \sim 1/\Gamma$ from the line of sight, while in the radio band, the region around $\theta \sim 0$ with low local PD is also bright\cite{granot99},
so that the net radio PD is lower\cite{shimoda21}.
As a result, the model polarization spectrum at 2.5 day (Figure \ref{polspec}; middle panel) is consistent with the upper limit on the radio PD. In this model, however, the polarization angle in the radio band is the same as that in the optical band. The difference in the observed PAs at the radio and optical bands does not support this model. 

The temporal changes of optical PD and PA in this model are plotted in Extended Data Figure \ref{lcall}. The PD changes as the angular size of visible region $\sim 1/\Gamma$ increases. It has the maximum value when $1/\Gamma \sim \theta_j + \theta_v$. The PA experiences a sudden 90 deg change at $t \sim 0.06\;$day, and it is constant before and after the time. The model with $\xi^2 = 0.56$ exhibits PD as high as $\simeq 5\%$ at $t \simeq 0.4\;$day, which is not consistent with the observed data. The less anisotropic turbulence leads to lower PD, as shown by the model with $\xi^2 = 0.81$ in Extended Data Figure \ref{lcall}, but it also appears incompatible with the observed data. 

\noindent {\bf Faraday depolarization model}\\
The low PD in the radio band could be ascribed to internal Faraday depolarization effect by cool electrons. The standard forward shock model usually assumes that all shocked electrons gain energy from shocked protons and form the power-law energy distribution $dn/d\gamma\propto\gamma^{-p}$ for $\gamma\gtrsim\gamma_{m} \sim \epsilon_{e}(m_{p}/m_{e})\Gamma$. Plasma particle simulations showed that all the electrons gain energy from shocked protons\cite{sironi11}, but it has not been confirmed by observations yet. Indeed, the forward shock model in which only a fraction $f (<1)$ of the total electrons is energized can also explain the observed afterglow light curves and spectra\cite{eichler05}.
In this case, the fraction $1-f$ of the total electrons remain as cool thermal electrons with Lorentz factor $\tilde{\gamma}_m = \eta\Gamma$ , where $\eta$ is a factor of the order of unity if the cool electrons are just isotropized at the shock front, and the correct physical parameters are
 $E_{\rm iso}'$ = $E_{\rm iso}/ f$, $n'=n/f$, $\epsilon_{e}'=\epsilon_{e}f$,
and $\epsilon_{B}'=\epsilon_{B}f$. 
The cool electrons cause Faraday depolarization on the synchrotron emission of the non-thermal electrons above self-absorption frequency\cite{toma08}.

We assume that the magnetic field in the shocked fluid is turbulent on hydrodynamic scale, which is comparable to the typical width of bright region in the shock downstream. Such field can be created by magnetohydrodynamic instabilities at the shock, such as Richtmyer-Meshkov instability\cite{sironi07, inoue13}.
For simplicity, we consider that the globally ordered magnetic field is negligible and that the plasma in the visible region consist of $N$ random cells in each of which magnetic field is ordered. At the optical band, for which the Faraday effect is not significant, the net PD is $P_{0} \sim \frac{(p+1)}{(p+7/3)}\frac{1}{\sqrt{N}}$, so that $N\sim5000$ can explain the optical PD $\sim$ 1\%. The Faraday rotation effect within the emission region results in the PD as\cite{sokoloff98} $P_{0}[(1-e^{-S})/S]$, where $S=(\nu/\tilde{\nu_{V}})^{-4}$
and $\tilde{\nu_{V}}$ is the frequency at which the Faraday depth is unity,
$\tilde{\nu}_V \sim 200 (1+z)^{-15/16}(\frac{1-f}{10f})^{1/2}\eta^{-1} \sqrt{\ln \tilde{\gamma}_m} N^{-1/12}(\frac{E_{\rm iso}}{10^{52} {\rm erg}})^{3/16} n^{9/16} (\frac{\epsilon_{B}}{0.01})^{1/4} (\frac{t}{1 {\rm day}})^{-1/16}$ GHz\cite{toma08}.
The middle panel of Figure \ref{polspec} shows the Faraday depolarization model for the radio and optical data, which indicates $\tilde{\nu}_V \gtrsim 100$ GHz. This leads to $f^{-1} - 1 \gtrsim 2.5\eta^{2}(\ln \tilde{\gamma}_m)^{-1}$.

\end{methods}

\begin{addendum}
 \item[Data Availability] Processed data are presented in the tables and figures in the paper. The ALMA data are available at ALMA Science Archive. The VLT data are available at ESO Science Archive Facility.

 \item[Code Availability] We used standard data reduction tools in Python and CASA \cite{casa}. The theoretical calculation code of flux and polarization used in this work is not publicly available. Results presented in this work are available from the corresponding author upon reasonable request.

 \item This work is based on observations collected at the European Southern Observatory under ESO programmes 0104.D-0600(C) and 0104.D-0600(A).This paper makes use of the following ALMA data: ADS/JAO.ALMA\# 2019.1.01016.T, 2019.1.01484.T. ALMA is a partnership of ESO (representing its member states), NSF (USA), and NINS (Japan), together with NRC (Canada), MOST and ASIAA (Taiwan), and KASI (Republic of Korea), in cooperation with the Republic of Chile. The Joint ALMA Observatory is operated by ESO, AUI/NRAO, and NAOJ. This work is supported by the Ministry of Science and Technology of Taiwan grants MOST 105-2112-M-008-013-MY3 (YU) and 106-2119-M-001-027 (KA). This work is also supported by JSPS Grants-in-Aid for Scientific Research No. 18H01245 (KT), No. 20J01086 (JS) and by Graduate Program on Physics for the Universe (GP-PU), Tohoku University (AK). KW acknowledges support through a UK Research and Innovation Future Leaders Fellowship awarded to dr. B. Simmons (MR/T044136/1), and support through an Alan Turing Institute Post-Doctoral Enrichment Award. We thank EA-ARC, especially Pei-Ying Hsieh for support in the ALMA observations.

 \item[Author Contribution] YU, KT, SC, and KW initiated the study. YU and KT mainly wrote the texts of this manuscript. YU, SC, KW, KYH, and ST managed the ToO observations and main data analysis. GP provided valuable thoughts for blushing up the ToO managements. KT, JS, AK, and SN managed the theoretical interpretations. AK played the principal role in numerical modeling. KA and HN checked the ALMA results. CC, KY, and MT provided $N_{H}$ analysis for evaluating intrinsic absorption using X- ray data. LI, JF, AUP, and MA provided VLT/X-shooter data. All of the authors contributed to the data analysis and discussed the results and the texts.
  
 \item[Competing Interests] The authors declare that they have no competing financial interests.
 
 \item[Correspondence] Correspondence and requests for materials should be addressed to Y.U.~(email: yjurata@gmail.com).

\end{addendum}

\clearpage


\begin{longtable}[c]{rcccccc}
\caption{{\bf Radio Polarization Observing Log.} Measurements with no special notation are summarized with 1-$\sigma$ errors.}
\label{almapol}
\\
\hline
Days & Frequency & PD & PA & I flux & Q flux  & U flux \\
     & [GHz]     & [\%]    & [deg] & [mJy] & [mJy] & [mJy] \\
\hline
0.476  & 97.5 & $<0.84$ (3-$\sigma$) & ---         & 3.587 $\pm$ 0.008 & --- & --- \\ 
1.458  & 97.5 & $<0.60$ (3-$\sigma$) & ---         & 4.912 $\pm$ 0.005 & --- & --- \\ 
2.525  & 97.5 & $<0.64$ (3-$\sigma$) & 37.7$-$52.3 &	3.949 $\pm$ 0.007 & 0.006 (rms) & $-0.023\pm0.006$ \\
5.482 & 97.5 & $<1.3$ (3-$\sigma$) & ---	& 2.371 $\pm$ 0.090 & 0.007 (rms) & 0.007 (rms) \\ \hline
\end{longtable}

\clearpage

\renewcommand{\tablename}{Extended Data Table}
\setcounter{table}{0}

\begin{longtable}[c]{rccc}
\caption{{\bf Summary of Optical Polarization}. Measurements with no special notation are summarized with 1-$\sigma$ errors. *The values for 0.121 and 0.417 days were derived in the wavelength range of $R$ band based on data reported by Buckley et al (2021)\cite{buckley21}.} 
\label{optpol}
\\
\hline
Days & Band & PD [\%] & PA [deg]  \\
\hline
0.121  & $R$* & $1.4\pm0.1$* & $68\pm5$*   \\ 
0.417  & $R$* & $1.0\pm0.1$* & $57\pm5$*  \\ 
2.525  & $R$                 & $1.3\pm0.1$                 & $61.6\pm6.3$ \\ 			
\hline 
\end{longtable}

\clearpage 

\begin{table*}[hbtp]
\caption{X-ray spectral properties for 8 known-$z$ GRBs with optical polarimetry}
\label{xray}
\centering
\begin{tabular}{ccccccc}
\hline
Events & Redshift & Galactic $N_{H}$ & Intrinsic $N_{H}$ & Spectral index ($\beta_{X}$) & Reduced  $\chi^{2}/$dof & A$_{V}$ \\
       &          & (10$^{20}$ cm$^{-2}$)         & (10$^{22}$ cm$^{-2}$) &               &                         & (mag) \\
\hline 
GRB080928  & 1.692 & 7.16 & 0.42$^{+0.16}_{-0.15}$ & $-1.07\pm0.07$ & 1.12/204 & 0.13 \\
GRB091018  & 0.971 & 3.07 & 0.35$^{+0.06}_{-0.05}$ & $-1.02\pm0.05$ & 0.98/266 & 0.11 \\
GRB091208B & 1.063 & 5.75  & 1.68$^{+0.29}_{-0.25}$ & $-1.13\pm0.09$ & 1.09/188 & 0.51 \\
GRB110205A & 2.22  & 1.70  & 0.56$^{+0.15}_{-0.14}$ & $-0.93\pm0.05$ & 0.97/264 & 0.17 \\
GRB121024A & 2.298 & 7.87  & 1.37$^{+0.38}_{-0.34}$ & $-1.00\pm0.08$ & 0.71/175 & 0.42 \\
GRB131030A & 1.293 & 5.62  & 0.59$^{+0.11}_{-0.10}$ & $-1.07\pm0.06$ & 0.97/244 & 0.18 \\
GRB190114C & 0.425  & 0.75 & 8.5$^{+0.3}_{-0.3}$  & $-1.09\pm0.06$ & 1.05/517 & 2.6  \\
GRB191221B & 1.148 & 8.62  & 0.16$^{+0.09}_{-0.08}$ & $-0.96\pm0.06$ & 0.88/246 & 0.05\\
\hline
\end{tabular}
\end{table*}

\clearpage

\begin{longtable}[c]{rcc}
\caption{{\bf Photometric Observing Log.} Measurements are summarized with 1-$\sigma$ errors and statistical errors only. *The optical flux at 2.385 days was derived using the acquisition images taken as part of the acquisition sequence of the VLT imaging polarimetry.}
\vspace{-0.7cm}
\label{tablephot}
\\
\hline
Days & Frequency [GHz] &  Flux density [mJy]\\
\hline
  1.523 & 97.5 & 4.579 $\pm$ 0.038  \\ 
  4.455 & 97.5 & 3.093 $\pm$ 0.042  \\ 
  8.375 & 97.5 & 1.320 $\pm$ 0.030  \\ 
  9.482 & 97.5 & 1.213 $\pm$ 0.016  \\
13.352 & 97.5 & 0.685 $\pm$ 0.035  \\ 
17.422 & 97.5 & 0.477 $\pm$ 0.053 \\ 
18.449 & 97.5 & 0.464 $\pm$ 0.017  \\
33.425 & 97.5 & 0.206 $\pm$ 0.039 \\ 
33.444 & 97.5 & 0.200 $\pm$ 0.019 \\ \hline
  1.471 & 145 &  5.823 $\pm$ 0.125   \\
  2.442 & 145 &  4.742 $\pm$ 0.099    \\
  8.444 & 145 &  1.046 $\pm$ 0.063    \\
  9.401 & 145 &  0.906 $\pm$ 0.022    \\
18.431 & 145 &  0.352 $\pm$ 0.018  \\ \hline
1.498 & 203 & 7.051 $\pm$ 0.589    \\
2.505 & 203 & 4.434 $\pm$ 0.205    \\ \hline 
2.385  & 4.59$\times10^{5}$  &0.027 $\pm$ 0.003*  \\ 			

\hline 
\end{longtable}

\clearpage

\clearpage

\clearpage
\renewcommand{\figurename}{Extended Data Figure}
\setcounter{figure}{0}

\begin{figure}
\centering
\includegraphics[width=15cm]{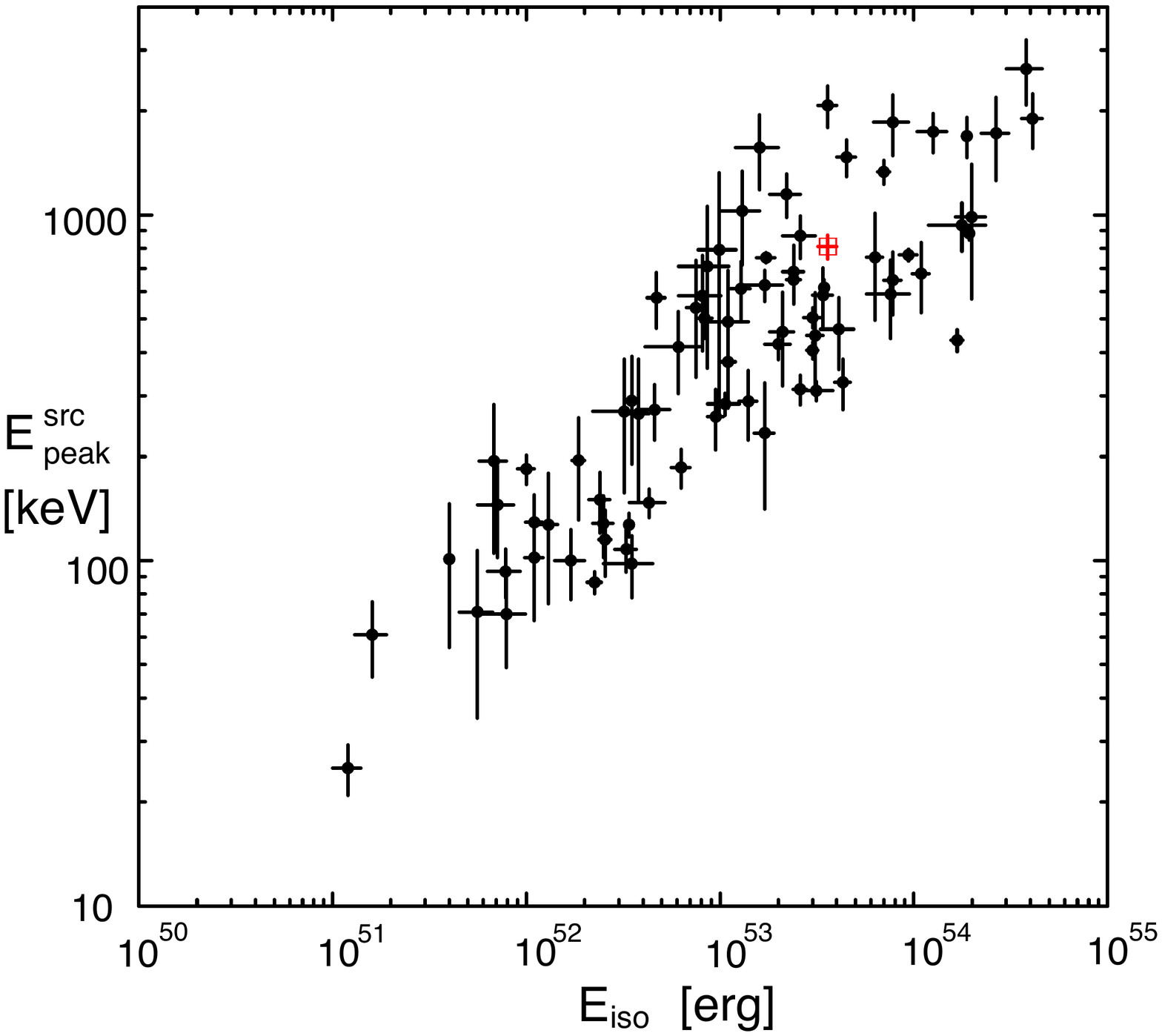}
\caption{{\bf $E^{\rm src}_{\rm peak}-E_{\gamma,{\rm iso}}$ relation \cite{amati02,yamaoka17}.} GRB 191221B, marked with the red box point, obeys the relation. All error bars represent 1-$\sigma$ uncertainties.}
\label{amati}
\end{figure}

\begin{figure}
\centering
\includegraphics[width=15cm]{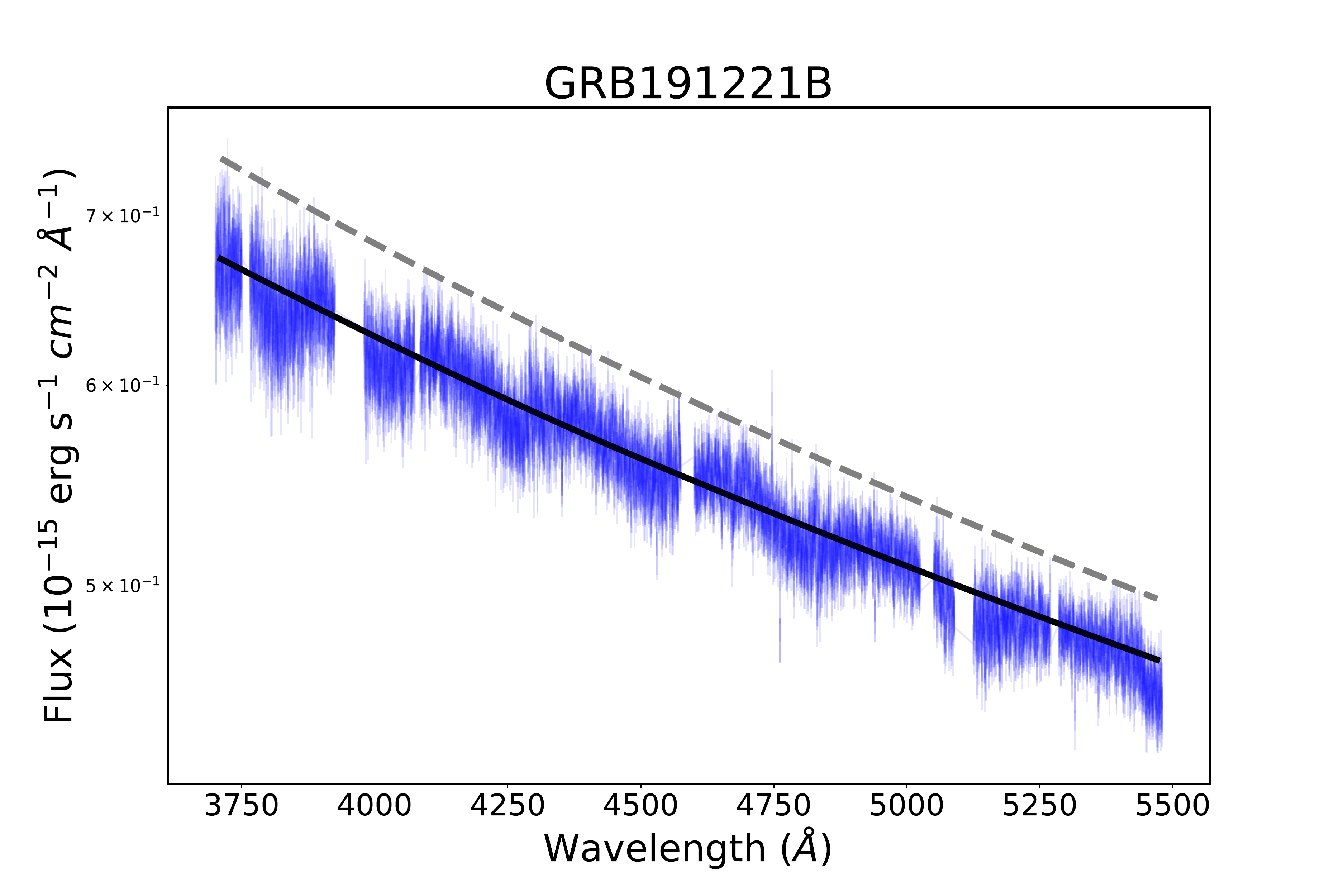}
\caption{{\bf Fit by a simple power-law and SMC \cite{gordon2003} extinction curve for the ultraviolet arm of the VLT-X-shooter. }
The black solid line is the best-fit and the dashed line shows the unestinguished afterglow spectrum. Marginalizing the power-law index and normalization, it turns out that the amount of rest-frame extinction is below 0.038\,mag (95\% upper limit).}
\label{xshooter}
\end{figure}

\begin{figure}
\centering
\includegraphics[width=15cm]{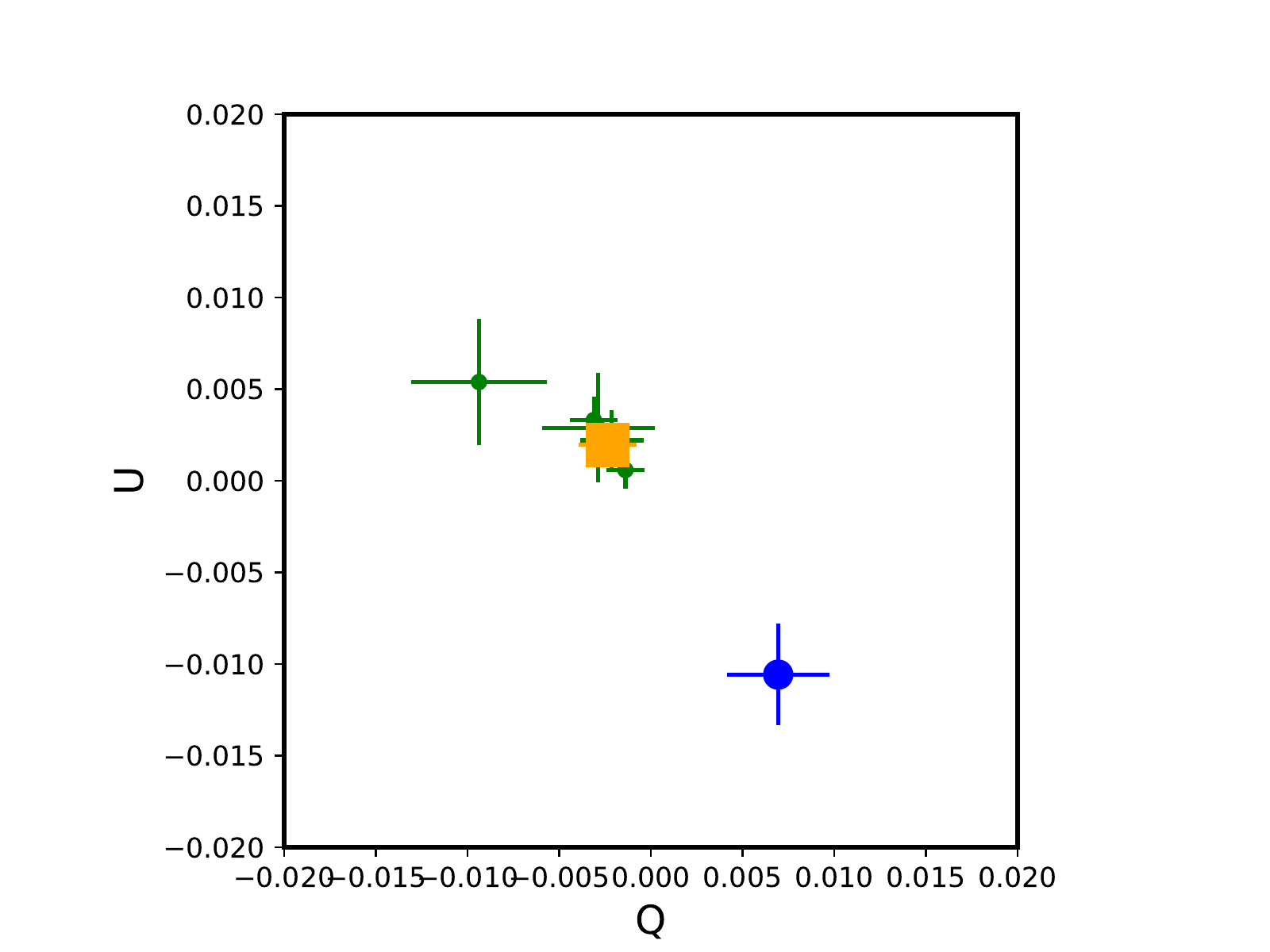}
\caption{{\bf Stokes raw Q and U measurements for field stars and the GRB afterglow. }
Green and blue circles indicate field stars and the afterglow, respectively.  The orange box shows the weighted average of the field stars.
}
\label{galactic}
\end{figure}

\begin{figure}
\centering
\includegraphics[width=9cm]{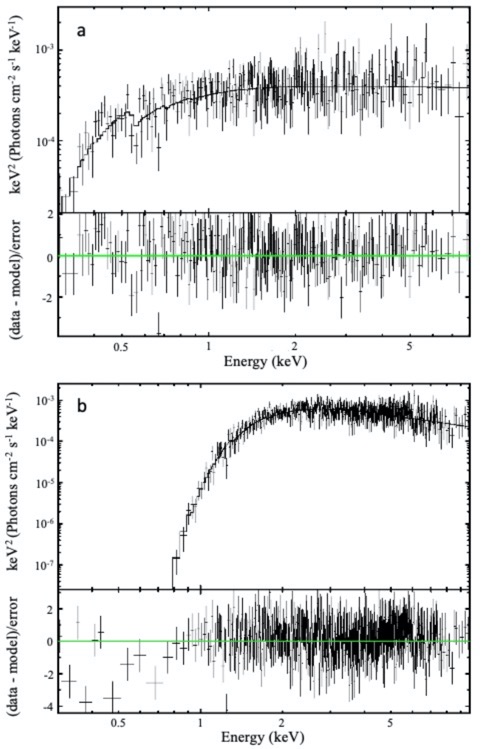}
\caption{{\bf X-ray spectrum for GRB191221B (a) and GRB190114C (b) afterglows.} a, GRB191221B X-ray afterglow spectrum described by a single power law modified with intrinsic and Galactic adsorptions, the latter of which is fixed at $N_{H}=8.6\times10^{20}$ ${\rm cm}^{-2}$. The derived best-fitting values of the intrinsic absorption column density and spectral index are $N_{H}=(1.6^{+0.9}_{-0.8})\times10^{21}$ ${\rm cm}^{-2}$ and $\beta_{X}=-0.96\pm0.06$, respectively, with reduced $\chi^{2}$/dof=0.88/246.
b, GRB190114C X-ray afterglow spectrum. The highly obscured spectrum with the intrinsic absorption column density of $N_{H}=(8.5^{+0.3}_{-0.3})\times10^{22}$ ${\rm cm}^{-2}$ is reasonable for the dust induced origin of polarization observed in the optical afterglow reported by Jordana-Mitjans et al. (2020) \cite{jordana-mitjans20}. All error bars represent 1-$\sigma$ uncertainties.}
\label{xspec}
\end{figure}

\begin{figure}
\centering
\includegraphics[width=8cm]{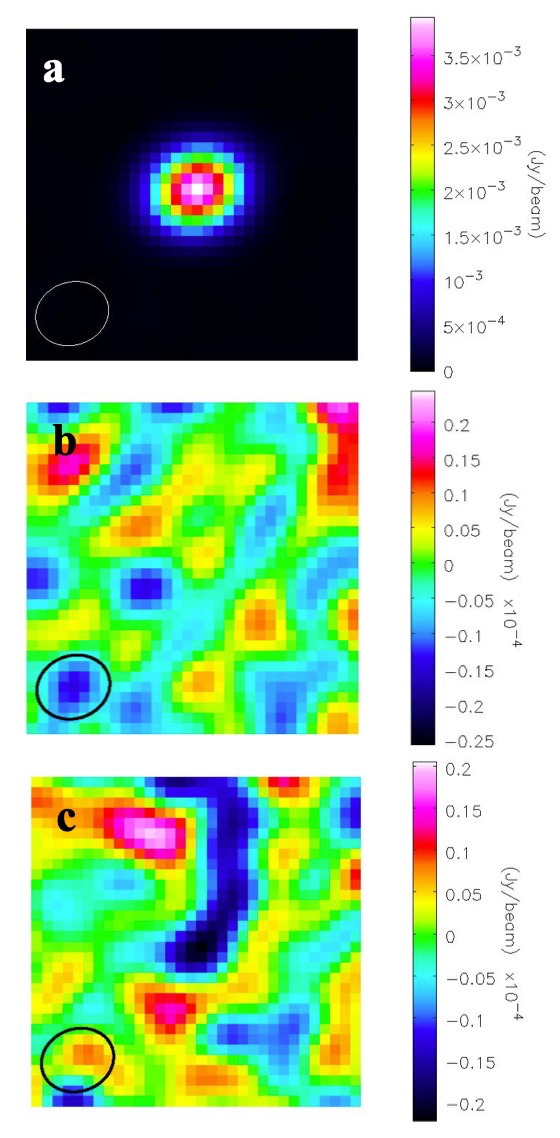}
\caption{{\bf Stokes $I$ (a), $Q$ (b), and $U$ (c) maps of the GRB191221B afterglow taken on 2019 December 24 (2.5 days after the burst). } The ALMA beam size is shown in the open circles. The map null detection on the Stokes-$Q$ map constrained the range of PA of 37.7-52.3 deg with 1$\sigma$ uncertainties.}
\label{stokesmap}
\end{figure}

\begin{figure}
\centering
\includegraphics[width=14cm]{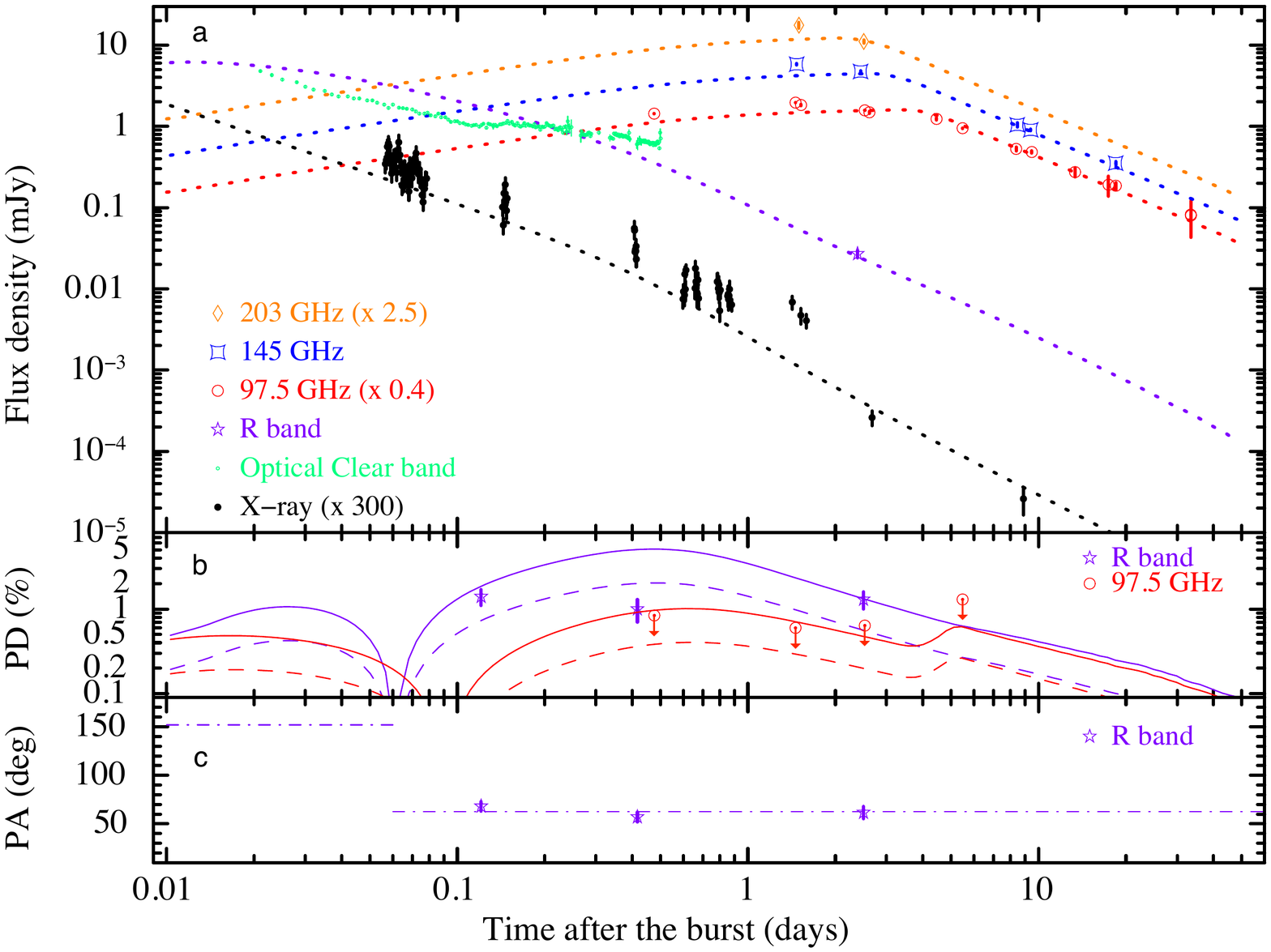}
\caption{{\bf GRB 191221B afterglow light curves in X-ray, optical, and radio bands together with optical and radio polarization variabilities.}  a, Multi-frequency light curves. Dotted lines indicate the model light curves fitted to radio, optical, and X-ray data observed by ALMA and VLT. Differences in the early optical afterglow (green small circles) and its wiggles may be caused by the magnitude-to-flux conversion of optical observations made by the very broad-band clear filter. 
b, PD temporal evolution in optical $R$ band (purple stars and lines) and radio 97.5 GHz band (3-$\sigma$ upper limits with red symbols and lines).  The solid and dashed lines indicate the plasma-scale turbulent magnetic field model with $\xi^2=0.56$ (solid) and $\xi^2=0.81$ (dashed). 
c, PA in the optical $R$ band.
The purple dashed-dot lines indicate the plasma-scale turbulent magnetic fields model with any $\xi^2$. All error bars represent 1-$\sigma$ uncertainties. The upper limits are at the 3-$\sigma$ level.}
\label{lcall}
\end{figure}


\clearpage

\section*{References}

\begin{thebibliography}{1}
\bibitem{gcn26534} Laha, S. et al.  GRB 191221B: Swift detection of a burst and a very bright optical candidate. \gcn  {\bf 26534} (2019)
\bibitem{gcn26537} Lipunov, V. et al. GRB 191221B: MASTER OT detection.  \gcn {\bf 26537} (2019)
\bibitem{buckley21} Buckley, D.~A.~H. et al. Spectropolarimetry and photometry of the early afterglow of the gamma-ray burst GRB 191221B. \mnras 506, 4621-4631 (2021)
\bibitem{gcn26553} Vielfaure, J.-B. et al.\  GRB 191221B: VLT/X-shooter redshift. \gcn {\bf 26553} (2019)
\bibitem{gcn26576} Frederiks, D. et al.\ Konus-Wind observation of GRB 191221B. \gcn {\bf 26576} (2019)
\bibitem{gcn26562} Sakamoto, T. et al.\ GRB 191221B: Swift-BAT refined analysis. \gcn {\bf 26562} (2019)
\bibitem{vltpolsys} Cikota, A. et al.\  Linear spectropolarimetry of polarimetric standard stars with VLT/FORS2. \mnras 464, 4146-4159 (2017) 
\bibitem{serkowski73} Serkowski, K.\ Interstellar Polarization (review). {\it IAUS}. 52, 145 (1973)
\bibitem{covino16} Covino, S. \& Gotz, D.\ Polarization of prompt and afterglow emission of Gamma-Ray Bursts. {\it Astron. Astrophys. Transactions} 29, 205-244 (2016)
\bibitem{urata19} Urata, Y. et al.\ First Detection of Radio Linear Polarization in a Gamma-Ray Burst Afterglow. \apjl 884, L58, 7 (2019)
\bibitem{sari98} Sari, R., Piran, T., \& Narayan, R.\ Spectra and Light Curves of Gamma-Ray Burst Afterglows. \apj 497, L17-L20 (1998)
\bibitem{zhang04} Zhang, B., M\'{e}sz\'{a}ros, P. Gamma-Ray Bursts: progress, problems \& prospects. {\it Int. J. Mod. Phys. A} 19, 2385-2472 (2004)
\bibitem{granot99} Granot, J., Piran, T., \& Sari, R.\ Images and Spectra from the Interior of a Relativistic Fireball. \apj513, 679-689 (1999)
\bibitem{shimoda21} Shimoda, J. \& Toma, K.\ Multi-wave band Synchrotron Polarization of Gamma-Ray Burst Afterglows. \apj 913, 58, 11 (2021)
\bibitem{panaitescu02} Panaitescu, A., \& Kumar, P.\ Fundamental Physical Parameters of Collimated Gamma-Ray Burst Afterglows. \apjl, 560, L49-L53 (2001).
\bibitem{rybicki79} Rybicki, G. B., \& Lightman, A. P. Radiative Processes in Astrophysics. {\it Wiley Int. Publ.} (1979)
\bibitem{medvedev99} Medvedev, M.~V. \& Loeb, A.\ Generation of Magnetic Fields in the Relativistic Shock of Gamma-Ray Burst Sources. \apj 526, 697-706 (1999)
\bibitem{sironi07} Sironi, L. \& Goodman, J.\ Production of Magnetic Energy by Macroscopic Turbulence in GRB Afterglows. \apj 671, 1858-1867 (2007)
\bibitem{gruzinov99} Gruzinov, A., \& Waxman, E. Gamma-Ray Burst Afterglow: Polarization and Analytic Light Curves. \apj 511, 852-861 (1999)
\bibitem{sagiv04} Sagiv, A., Waxman E., \& Loeb, A. Probing the Magnetic Field Structure in Gamma-Ray Bursts through Dispersive Plasma Effects on the Afterglow Polarization. \apj 615, 366-377 (2004)
\bibitem{toma08} Toma, K., Ioka, K., \& Nakamura, T.\ Probing the Efficiency of Electron-Proton Coupling in Relativistic Collisionless Shocks through the Radio Polarimetry of Gamma-Ray Burst Afterglows.  \apj 673, L123-L126 (2008)
\bibitem{spitkovsky08} Spitkovsky, A. Particle Acceleration in Relativistic Collisionless Shocks: Fermi Process at Last?. \apj 682, L5-L8 (2008)
\bibitem{keshet09} Keshet, U., Katz, B., Spitkovky, A., \& Waxman, E. Magnetic Field Evolution in Relativistic Unmagnetized Collisionless Shocks. \apj 693, L127-L130 (2009)
\bibitem{sari99} Sari, R.\ Linear Polarization and Proper Motion in the Afterglow of Beamed Gamma-Ray Bursts. \apj, 524, L43-L46 (1999)
\bibitem{ghisellini99} Ghisellini, G., \& Lazzati, D. Polarization light curves and position angle variation of beamed gamma-ray bursts. \mnras 309, L7-L11 (1999)
\bibitem{inoue13} Inoue, T., Shimoda, J., Ohira, Y., et al.\ The Origin of Radially Aligned Magnetic Fields in Young Supernova Remnants. \apjl 772, L20, 5 (2013)
\bibitem{eichler05} Eichler, D. \& Waxman, E.\ The Efficiency of Electron Acceleration in Collisionless Shocks and Gamma-Ray Burst Energetics. \apj 627, 861-867 (2005)
\bibitem{sokoloff98} Sokoloff, D.~D., Bykov, A.~A., Shukurov, A., et al.\ Depolarization and Faraday effects in galaxies \mnras 299, 189-206 (1998)
\bibitem{murase06} Murase, K., Ioka, K., Nagataki, S., et al.\ High-Energy Neutrinos and Cosmic Rays from Low-Luminosity Gamma-Ray Bursts?. \apj 651, L5-L8 (2006)
\bibitem{kimura17} Kimura, S.~S. et al.\ High-energy Neutrino Emission from Short Gamma-Ray Bursts: Prospects for Coincident Detection with Gravitational Waves. \apjl 848, L4, 6 (2017).
\bibitem{casa} McMullin, J.~P. et al.\ CASA Architecture and Applications. ASP Conf. Series 376, 127 (2007)
\bibitem{goldoni2006} Goldoni, P. et al.\ Data reduction software of the X-shooter spectrograph. SPIE Conf. Series 6269, 2 (2006)
\bibitem{modigliani2010} Modigliani, A. et al.\ The X-shooter pipeline. SPIE Conf. Series 7737, 28 (2010)
\bibitem{selsing2019} Selsing, J. et al.\  The X-shooter GRB afterglow legacy sample (XS-GRB). \aap 623, 92, 42 (2019)
\bibitem{carnal2017} Carnall, A.C.\ SpectRes: A Fast Spectral Resampling Tool in Python. arXiv:1705.05165 (2017) 
\bibitem{gordon2003} Gordon, K.D. et al.\ A Quantitative Comparison of the Small Magellanic Cloud, Large Magellanic Cloud, and Milky Way Ultraviolet to Near-Infrared Extinction Curves. \apj 594, 279-293 (2003)
\bibitem{gordon2009} Gordon, K.D. et al.\ FUSE Measurements of Far-Ultraviolet Extinction. III. The Dependence on R(V) and Discrete Feature Limits from 75 Galactic Sightlines. \apj 705, 1320-1335 (2009)
\bibitem{covino2013} Covino, S. et al.\ Dust extinctions for an unbiased sample of gamma-ray burst afterglows. \mnras 432, 1231-1244 (2013)
\bibitem{patat2015} Patat, F. et al.\ Properties of extragalactic dust inferred from linear polarimetry of Type Ia Supernovae. \aap 557, 53, 10 (2015)
\bibitem{covino03} Covino, S., Malesani, D., Ghisellini, G., et al.\ Polarization evolution of the GRB 020405 afterglow. \aap 400, L9-L12 (2003)
\bibitem{wiersema12} Wiersema, K., Curran, P.~A., Kr{\"u}hler, T., et al.\ Detailed optical and near-infrared polarimetry, spectroscopy and broad-band photometry of the afterglow of GRB 091018: polarization evolution. \mnras 426, 2-22 (2012)
\bibitem{uehara12} Uehara, T., Toma, K., Kawabata, K.~S., et al.\ GRB 091208B: First Detection of the Optical Polarization in Early Forward Shock Emission of a Gamma-Ray Burst Afterglow. \apjl 752, L6, 5 (2012)
\bibitem{gorosabel11} Gorosabel, J. et al.\ GRB 110205A: detection of optical linear polarization from CAHA. \gcn {\bf 11696} (2011)
\bibitem{wiersema14} Wiersema, K. et al.\ Circular polarization in the optical afterglow of GRB 121024A. \nat 509, 201-204 (2014)
\bibitem{king14} King, O.~G. et al.\ Early-time polarized optical light curve of GRB 131030A. \mnras 445, L114-L118 (2014)
\bibitem{jordana-mitjans20} Jordana-Mitjans, N. et al.\ Lowly Polarized Light from a Highly Magnetized Jet of GRB 190114C. \apj 892, 97, 17 (2020)
\bibitem{evans07} Evans, P.~A. et al.\ An online repository of Swift/XRT light curves of $\gamma$-ray bursts. \aap 469, 379-385 (2007)
\bibitem{evans09} Evans, P.~A. et al.\ Methods and results of an automatic analysis of a complete sample of Swift-XRT observations of GRBs. \mnras 397, 1177-1201 (2009)
\bibitem{willingale13} Willingale, R. et al.\ Calibration of X-ray absorption in our Galaxy. \mnras 431, 394-404 (2013)
\bibitem{wilms00} Wilms, J., Allen, A., \& McCray, R.\ On the Absorption of X-Rays in the Interstellar Medium. \apj 542, 914-924
\bibitem{schady10} Schady, P. et al.\ Dust and metal column densities in gamma-ray burst host galaxies. \mnras 401, 2773-2792 (2010)
\bibitem{zhang01} Zhang, B., \& M\'{e}sz\'{a}ros, P. Gamma-Ray Burst Afterglow with Continuous Energy Injection: Signature of a Highly Magnetized Millisecond Pulsar. \apj 552, L35-L38 (2001)
\bibitem{rees98} Rees, M. J., \& M\'{e}sz\'{a}ros, P. Refreshed Shocks and Afterglow Longevity in Gamma-Ray Bursts. \apj 496, L1-L4 (1998)
\bibitem{zhang09} Zhang, W. \& MacFadyen, A.\ The Dynamics and Afterglow Radiation of Gamma-Ray Bursts. I. Constant Density Medium. \apj 698, 1261-1272 (2009)
\bibitem{vaneerten12} van Eerten, H.~J. \& MacFadyen, A.~I.\ Observational Implications of Gamma-Ray Burst Afterglow Jet Simulations and Numerical Light Curve Calculations. \apj 751, 155 (2012)
\bibitem{gao13} Gao, H., et al.\ A complete reference of the analytical synchrotron external shock models of gamma-ray bursts. New Astron. Rev. 57, 141 (2013)
\bibitem{sari00} Sari, R. \& M\'{e}sz\'{a}ros, P.\ Impulsive and Varying Injection in Gamma-Ray Burst Afterglows. \apj 535, L33-L37 (2000). 
\bibitem{yost03} Yost, S. A., Harrison, F. A., Sari, R., \& Frail, D. A.\ A Study of the Afterglows of Four Gamma-Ray Bursts: Constraining the Explosion and Fireball Model. \apj, 597, 459-473 (2003).
\bibitem{cenko10} Cenko, S. B., et al.\ The Collimation and Energetics of the Brightest Swift Gamma-Ray Bursts. \apj 711, 641-654 (2010).
\bibitem{rossi04} Rossi, E. M., Lazzati, D., Salmonson, J. D., \& Ghisellini, G. The polarization of afterglow emission reveals $\gamma$-ray bursts jet structure. \mnras 354, 86-100 (2004)
\bibitem{sironi11} Sironi, L. \& Spitkovsky, A.\ Particle Acceleration in Relativistic Magnetized Collisionless Electron-Ion Shocks. \apj 726, 75 (2011) 
\bibitem{amati02} Amati, L. et al.\ Intrinsic spectra and energetics of BeppoSAX Gamma-Ray Bursts with known redshifts. \aap 390, 81-89 (2002) 
\bibitem{yamaoka17} Yamaoka, K. et al.\ Suzaku Wide-band All-sky Monitor (WAM) observations of GRBs and SGRs. \pasj 69, R2 (2017) 

\end{thebibliography}
\end{document}